\documentclass[manuscript,nonacm]{acmart}

\renewcommand\footnotetextcopyrightpermission[1]{} 
\setcopyright{none}

\usepackage{float}
\usepackage{tabularx}
\usepackage{multirow}
\usepackage{pdfpages}
\newcolumntype{L}{>{\raggedright\arraybackslash}X}
\usepackage{booktabs}
\usepackage{tikz}
\usetikzlibrary{arrows.meta,positioning,shapes,fit}

\usepackage{xcolor}

\AtBeginDocument{%
  }

\begin{document}
\title{Terms of Care: Designing for Participation in Data Governance by Disability Communities}

\author{Madiha Zahrah Choksi}
\affiliation{%
  \institution{Columbia University}
  \city{New York}
  \country{United States}}
\email{m.choksi@columbia.edu}

\author{Reeda Shimaz Huda}
\affiliation{%
  \institution{Georgia Institute of Technology}
  \city{Atlanta}
  \country{United States}
}
\email{reedahuda@gmail.com}

\author{Rita Faia Marques}
\affiliation{%
  \institution{Microsoft Research}
  \city{Cambridge}
  \country{United Kingdom}
}
\email{t-ritaf@microsoft.com}

\author{Anja Thieme}
\affiliation{%
  \institution{Microsoft Research}
  \city{Cambridge}
  \country{United Kingdom}
}
\email{anthie@microsoft.com}

\author{Cecily Morrison}
\affiliation{%
  \institution{Microsoft Research}
  \city{Boston}
  \country{United States}
}
\email{cecilym@microsoft.com}

\renewcommand{\shortauthors}{Choksi et al.}

\begin{abstract}
Improving disability representation in AI-generated media requires community-specific datasets, yet
openly sharing such data can expose communities to misuse, decontextualization, and exploitation. We
examine how disability communities can participate in dataset governance through context-sensitive
licensing. Working with three disability advocacy organizations, we conducted semi-structured interviews,
an educational webinar and modular license activity to promote community participation in data
stewardship. Our findings show that governance in these high-context communities relies on organizational
memory (especially situated harms); both formal authority and informal relational practices;
communication; and resulting trust. The modular license activity made organizational data choices more
accessible and values-led, allowing communities to adapt Creative Commons clauses to express permitted
uses, sector-specific restrictions, safeguarding requirements, and commitments to respectful
representation. Yet, concerns remain about monitoring and enforcing appropriate data use. We contribute
(1) an empirical account of governance tensions, (2) a practical modular approach, and (3) design
implications for AI data platforms that support relational, context-sensitive governance beyond
standardized, one-size-fits-all frameworks.
\end{abstract}




\maketitle

\section{Introduction}
Previous research has shown that AI media generation often poorly represents disability communities~\cite{gadiraju2023wouldn, mack2024they, bennett2025toward, sadeghiani2024generative}, magnifying existing historical biases and creating material barriers to education, employment, and livelihoods~\cite{whittaker19disability}. 
Simultaneously, the speed and scale at which AI media generation is being adopted offers an opportunity to significantly reshape public perception of disability by producing generated media that more accurately reflects lived experiences~\cite{genaichi,wagner2019word}. 
AI data stewardship that defines 'good representation' and encodes it in training data is one potential avenue for improving disability representation in AI-generated media, and consequently in public perception.
Communities, and the advocacy organizations that amplify community voice, must balance how data provides opportunities for representation and advocacy with the risks of that data being exploited or misused. 
There is a documented history in which people with disabilities have seen their images circulated without permission, whether framed as objects of pity, ridiculed and harassed online, or appropriated without credit in creative and technological projects~\cite{itscomplicated, alongthemargins}. The use of AI image generators to create fictitious influencers with Down syndrome to sell sexual content is one recent example~\cite{maiberg2025ai, hall2025human, jo2020lessons}, and it is not isolated. Platforms have been found hosting AI-generated accounts that sexualize and impersonate people with disabilities, including those with Down syndrome, vitiligo, and albinism~\cite{evans2026instagram}. Community data stewards must work towards safeguarding community-specific data by negotiating and articulating the terms of their use.

While there are a range of data governance mechanisms that can be brought to bear, for those organizations that choose to open-source via a platform such as GitHub\footnote{\href{https://docs.github.com/en/repositories/managing-your-repositorys-settings-and-features/customizing-your-repository/licensing-a-repository}{GitHub: Managing your repository}}, Hugging Face\footnote{\href{https://huggingface.co/docs/hub/en/repositories-licenses}{Hugging Face: Repository Licenses}} or Kaggle\footnote{\href{https://www.kaggle.com/discussions/getting-started/116476}{Kaggle: Common license types for datasets}}, the core mechanism is data licensing~\cite{bustamante2018participatory,contractor2022behavioral}. Open-source licenses can make concrete the protections and boundaries that communities need, enabling external actors to engage with datasets in responsible and transparent ways. Yet, most existing licensing frameworks for open-source AI models and datasets are designed as standardized tools for broad reuse. Common examples include: (i) the BigScience Large Open-science Open-access Multilingual Language Model (BLOOM) -- a collaborative open-source large language model project; (ii) the Responsible AI Licenses (RAIL) -- a set of licenses imposing usage restrictions to promote responsible AI; and (iii) the Copyleft-next -- a modernized copyleft license. These approaches, however, rarely account for the relational qualities and values that shape decision-making within disability communities. In such high-context cultures~\cite{hall1973silent}, which are defined by their shared histories, trust, long-standing relationships, and a complex set of implicit norms underpinning governance, current licensing approaches risk flattening the nuance of community priorities, potentially reinforcing inequities. 

Despite a growing body of work on open source AI governance and AI ethics~\cite{bigham10VizWiz,choksi2024licenses,delacroix2019bottom,mackkelly,viljoen2021relational}, there remains a gap in scholarship on effective data licensing approaches for high-context communities. This paper builds on prior research that sought to improve disability representation in AI image generation models by actively involving disability communities in AI data practices~\cite{6858}. In that work, researchers collaborated with three disability advocacy organizations representing people with dwarfism and people with vision impairments to develop an early technical prototype of a community-centered AI data stewardship platform~\cite{6858}. 
The platform is designed to support communities to collectively define what ‘good' representation means to them, and to curate a community-centered image dataset. Each dataset comprises 400 photos showing a wide range of community members across different activities and settings, together with structured annotations and metadata formatted specifically for AI training and evaluation. 

While prior work focused on supporting community participation in dataset creation for AI, questions regarding the governance and licensing of these datasets remained unresolved. This paper addresses this gap through the design and usage of a \textit{modular license} that supports high-context communities in defining appropriate data licensing terms for their image datasets -- drawing on the same project context and collaborating organizations. We first offer an empirical account of governance norms and tensions of the three communities that we work with, showing how histories of harm and resource constraints shape data stewardship practices. We use this to motivate a modular license design, based on a Creative Commons core, that scaffolds organizations inscribing their values into a data license. 
Finally, we evaluate the experience of three disability communities in creating bespoke data licenses. Our contributions are:
\begin{enumerate}
    \item An empirical account of governance tensions faced by high-context communities~\cite{hall1973silent}, highlighting the relationship between informal and formal practices in negotiations of data use and the formation of trust.
    \item A practical template enabling high-context communities to define a meaningful open-source license for their own data that balances the benefits of providing data for AI with protection against misuse and harm.
    \item Concrete design implications for AI platforms on how they can configure community participation in developing licenses that are both usable and contextually meaningful.
\end{enumerate}
\section{Related Work}
We situate our work within four areas of literature: (1) disability representation in AI, (2) high-contexts communities, (3) data governance, and (4) open-source licensing.

\subsection{Disability Data Representation and AI}
Recent work in HCI has illustrated how AI image generation systems can be disrespectful and misaligned to the lived experience of people with disabilities~\cite{glazko2023auto, mack2024they}. Indeed, this is reflected across all types of AI systems, such that these systems work less well for people with disabilities~\cite{heung25ignorance,kane17at,massiceti2024explainingclipsperformancedisparities,venkit23automated}, or that they are negatively stereotyped or erased in generated media~\cite{gadiraju2023wouldn, mack2024they}. To address these challenges, a number of efforts have been made to work with communities to collect community-specific data. Key examples include: ORBIT~\cite{orbit2021} - videos and images of objects recorded by people who are blind or low-vision; StammerTalk~\cite{publicizemystutter} by people with atypical speech; and ASL Citizen~\cite{desai2023aslcitizencommunitysourceddataset} by people who are Deaf/Hard of hearing. In addition, we see more grassroots approaches like the Disabled And Here stock image collection celebrating disabled Black, Indigenous, People of Colour (BIPOC)~\cite{disabledandhere} and multi-organization supported initiatives like the Disability:IN Corporate Disability Stock Photography Collection~\cite{disabilityin_stockphotography}. 

Although some advocates have called for the inclusion of more disability-related data, these strategies can introduce new risks for harm, including heightened surveillance, privacy violations, and the reinforcement of narrow definitions of disability~\cite{whittaker19disability}. Mankoff et al.~\cite{mankoffjen} called for researchers to adopt a critical lens informed by disability studies, ensuring that broader issues of access, representation, and the autonomy of disability communities over technologies impacting their lives are not sidelined~\cite{mackkelly}. We adopt the perspective that ``disability'' is a political and social identity~\cite{kafer25feminist} that should inform \textit{how AI systems are designed, what protections are needed, and who has the right to define acceptable uses of data}.

\subsection{High-context Communities}
We draw on anthropologist Edward T. Hall's concept of high-context cultures~\cite{hall1973silent, hall1976beyond} to characterize the disability advocacy communities we engage with. In high-context settings, communication and decision-making rely on shared histories, trust, long-standing relationships, and a dense set of implicit norms. Members understand and negotiate protocols through unspoken cues and collective memory more than through formal rules. Importantly, this stands in contrast to the more formal mechanisms of governance, which assume that obligations can be explicitly articulated and enforced through documented systems.

We invoke high-context not as a fixed cultural type but as a situated description of how these communities actually coordinate, a distinction that matters given critiques of Hall's concept when it is used to rank national cultures on thin empirical grounds~\cite{kittler2011special}. This framing also reveals a central friction. The data governance tools available, from consent notices and licenses to enforcement regimes, assume a low-context mode of explicit articulation of norms and obligations, whereas high-context communities often govern through implicit, relational practices. This paper makes a novel contribution by examining this mismatch and showing \textit{how licensing can be redesigned to translate community-specific norms into explicit, enforceable terms without stripping them of their relational context}. 


\subsection{Data Governance}
The above described tension between implicit community norms and explicit governance mechanisms is visible in the literature on data governance, which documents significant challenges with common approaches for people from historically marginalized communities. Existing governance tools tend to emphasize individual responsibility, particularly through privacy notices and consent agreements~\cite{simms}.
This places the burden of informed decision-making on individuals, many of whom lack the resources or expertise to understand how their data will be used, which can create undue risks and contribute to downstream harms~\cite{RENAUD2018198,simms}. 
In AI pipelines specifically, core governance mechanisms such as consent, provenance, and transparency are routinely undermined as datasets are scraped or aggregated without consultation~\cite{longpre2024consent}. 

To contrast this focus on individual rights, scholars have increasingly turned toward collective governance models and relational approaches~\cite{dafoe2018ai}. Viljoen~\cite{viljoen2021relational}, for example, proposes a \textit{relational theory of data governance}, and shows how data practices are embedded within social relationships and institutional structures, not only individual data subjects. 

Several participatory governance models have been designed to redistribute power and strengthen community agency in decisions about data. Data trusts invite a trustee to hold and govern data on behalf of beneficiaries and is obligated to act in their best interests~\cite{delacroix2019bottom,mendonca2025datacooperativesdemocraticmodels}. 
Although promising, data trusts face substantial legal and practical challenges. For example, key GDPR rights such as portability and erasure currently cannot be transferred to a third-party trustee~\cite{delacroix2019bottom}. Infrastructural constraints exacerbate these issues, as Olorunju and Adams~\cite{olorunju2024african} document in South Africa. Communities often rely on customary processes because formal data protection authorities are under-resourced, raising concerns that data trusts may reinforce existing power asymmetries rather than mitigate them. 
A complementary approach is the data cooperative. Mendonca et al.~\cite{mendonca2025datacooperativesdemocraticmodels} describe cooperatives as member-driven, democratic organizations in which individuals pool their data and collectively determine governance, access, and benefit sharing. Rooted in collaborative governance principles~\cite{guttmann2021commons,vollan2010cooperation}, data cooperatives offer communities a mechanism for more direct and collective control over downstream use. 

Building on these institutional models, researchers in HCI and critical data studies similarly advocate for collective and participatory approaches to data stewardship~\cite{sociotechnicalaigov,jakestein,bogen}. 
Kamikubo et al.~\cite{kamikubo} argue that meaningful governance requires moving beyond transparency toward participatory data stewardship that foregrounds contributors in dialogue and decision-making. However, in the context of AI, participatory approaches often fall short because community-based methods are designed for localized contexts, while AI systems operate at global scale. Young et al.~\cite{young2024participation} call for infrastructures that enable communities to designate representatives and participate in governance in ways aligned with existing power structures and responsibilities. This paper addresses this challenge by examining \textit{how high-context communities can translate relational norms and collective priorities into concrete, enforceable licensing terms, thereby making community participation actionable within data-governance infrastructure}. 

\subsection{Open-Source Licensing}
Not all governance mechanisms require the creation of new institutional bodies. Another alternative is governance tools that are adaptable and deployable within existing organizational structures~\cite{choksi2024licenses,lessig2004creative}. 
Open-source licensing frameworks are such an example and could be seen as a tractable entry point for community-centered data governance. Copyright law establishes default rules granting creators exclusive rights to copy, distribute, modify, and adapt their works through the use of licenses. Scholars show that open-source licenses function as more than legal instruments, but as community-authored governance artifacts that articulate shared norms, values, and boundaries~\cite{choksi2024licenses,kelty2020two}. This view resonates with work on ``data leverage''~\cite{NICK2,nick3,nick1}, which positions licensing as a means for contributors to negotiate the conditions under which their data may be accessed or reused. Our approach aligns with recent calls for participatory and relational forms of data stewardship~\cite{NICK2,nick3,nick1} through a focus on licensing, while acknowledging that it is only one component of a larger governance ecosystem. 

Research and practice in ethical AI licensing similarly demonstrates how licenses can encode forms of governance, such as use purpose limitations and culturally specific constraints~\cite{contractor2022behavioral,longpre2024consent} that are difficult to express through technical metadata. Prior work distinguishes licenses that emphasize broad reuse (e.g., Creative Commons, Apache, MIT), and specialized licenses that introduce use-restrictions, obligations, or contextual constraints (e.g., GPL Open RAIL)~\cite{coleman2013coding,contractor2022behavioral,lessig2004creative}. Broad licenses provide space to articulate community needs in a semantically rich way, while more specialized licenses determine which downstream uses are permitted at the time of release. The latter requires enough knowledge to have foresight about how things might go wrong, while the former relies on the goodwill of those using the data to respect its intended use. We see an important function of open-source licenses as a ‘boundary object’~\cite{chen2004identifying, lee2007boundary} that allows disability organizations to articulate governance priorities and values, and concretize concerns about downstream use. 

This paper extends the current opportunity to choose between these two types of licenses by providing disability organizations opportunity to selectively modify and combine licensing clauses to express their relational governance needs. We build off the Creative Commons open-source licensing model which is often used for visual media. Platforms such as Flickr, Wikimedia Commons, and YouTube have integrated CC licensing, allowing creators to mark works in ways that are legally meaningful and practically visible~\cite{lessig2004creative}. 
The types of clauses these licenses might include are: attribution requirements (BY), limits on commercial use (NC), prohibitions on derivative works (ND), and requirements to share derivative works under the same terms (SA). These components enable a spectrum of openness instead of a binary choice between ``all rights reserved'' copyright and the public domain. As such, this paper seeks to demonstrate \textit{how participatory licensing can provide a practical governance infrastructure through which high-context communities collectively define, formalize, and communicate binding conditions for the downstream use of their data}.

\section{Study Method \& Modular License Design}
The development and evaluation of community-specified modular licenses was carried out in two phases. First, semi-structured interviews were conducted with each participating disability advocacy organization to understand the governance structures and decision-making processes of these high-context communities. Insights from these interviews then informed the design of the modular license activity, which was subsequently used and evaluated with the participating communities.

\subsection{Community Participation}
This study draws on the project context and collaborating organizations described in prior work~\cite{6858}. 
The organizations were recruited by an international NGO, with a focus on organizations that represented \textit{people with dwarfism and people with vision impairments.}\footnote{A manual, human review by two disability experts of images generated for 662 prompts (612 benign, 50 adversarial), prompted across ten main types of disabilities (e.g., mental health, blindness, low vision, mobility, neurodiversity, learning, speech, deafness, hard of hearing, pan disability) using an off-the-shelf image generation service found that defect rates were higher for Dwarfism and Blind and Low Vision communities.} 
Each organization signed a contract that included financial support for giving time and focus to the research. 
Concretely, the organizations were asked to partake in research activities; curate a dataset of 400 images with corresponding annotations; and open-source the final dataset for greater reach and impact. 

For each organization, the primary participant in the data licensing process was the project lead. Within the study context, project leads took on the role as data stewards for their organization's membership. They were the main liaison between the research team and their communities and responsible for all final decisions about dataset structure and content. They also sourced data for their image datasets and they took on the critical role of helping their community form a viewpoint on AI representation. Two project leads were themselves members of the disability community they represented. The study was reviewed for ethics and IRB approved. Informed consent was sought from all participants in writing prior to the study.

\subsection{Phase 1: Organizational Governance Interviews}
We began our study by conducting semi-structured interviews with each organizations' project lead -- referred to as P1, P2 and P3 -- independently. Through these interviews, our goals were to map established governance practices and identify tensions and trade-offs that organizations must navigate. 
The protocol (Appendix: \autoref{appendix:interview-protocol}) covered four broad areas: (1) guiding principles that shaped the organizations' approaches to data stewardship, community representation, and trust building; (2) current organizational data practices for consent and sharing member data; (3) organizational experiences of conflict over information sharing; and (4) explicit governance tools and roles (e.g., board votes). 

Each interview lasted approximately 90 minutes. The interviewer guided the discussion to understand in depth the ways that formal and informal structures (i.e., cultural norms) governed decision-making processes, including those specific to data sharing and technology; the emergence of new policies or norms; and not least members' expectations for privacy. Throughout, project leads were invited to ask questions and express any other concerns regarding privacy, consent, or data governance that could be addressed in the subsequent parts of our study. These insights were used to contextualize the choice of modular clauses through relevant examples in the Phase 2 activity. 

All interviews were video-recorded and transcribed. Additionally, two researchers maintained detailed field notes documenting emergent themes and explicit governance scenarios shared by participants. We used qualitative content analysis~\cite{forman2007qualitative} to examine project leads' experiences and address our more practical design questions. This approach is well suited to our evolving research process as it supports concurrent data collection and analysis, comparison within and across communities, and the integration of different data formats like transcripts, research notes, and survey responses~\cite{forman2007qualitative, krippendorff1996content}. All authors were involved in the data analysis that included the steps of: (1) data immersion through reflective notes after each activity; (2) data coding whereby research findings were deductively categorized based on research phase and activity, while allowing themes to emerge inductively; and (3) interpretation through descriptive summaries and narrative analysis as documented through our finding sections.

\subsection{Phase 2: Modular License Design \& Evaluation}
Building on Phase 1's new understandings of how informal and formal governance practices play together in high-context communities, Phase 2 examined how these governance logics materialized in a modular license design activity, where participant selectively adapt the Creative Commons license to encode their political commitments, red lines, and community values -- an activity that first is supported by a data governance webinar.

\subsubsection{Data Governance Webinar}
To set the context for designing a license by each organization, we began with an interactive educational data governance webinar to introduce key concepts in copyright and open-source AI data governance. This 90-minute educational session, led by the first author, aimed to translate abstract legal and technical concepts for a lay audience. All participants attended the webinar as a group to support group reflection and peer exchange. The webinar included five broad education topic areas: (1) copyright and licensing basics, including the distinction between owning a work and granting permission to others; (2) different types of open-source licenses, explaining the differences between copyleft and permissive licenses using common examples of GNU GPL, Creative Commons Share-Alike (CC BY-SA), MIT, and Apache licenses; (3) key license terms such as derivative works, attribution, non-commercial clauses, no-harm clauses, share-alike requirements, and ethical use provisions; (4) the importance of datasheets or dataset cards that serve as ``nutrition labels'' to support data transparency and provenance: who collected the data, why, and under what conditions it can be used~\cite{gebru2021datasheets, mitchell2019model, alkemade2023datasheets}; and (5) an introduction to Hugging Face as one of the leading real-world platforms for hosting open source datasets, highlighting specifically platform affordances for license documentation, access gating and consent, and notices for the removal of sensitive materials. The full webinar deck is available in the Supplementary Materials. 

Further, to ground the many abstract legal concepts in participants' lived organizational experiences, the webinar was interweaved with four interactive components: 
\begin{enumerate}
    \item \textit{Building a Community Library}: Participants were asked to imagine their dataset as a traditional `library collection' and asked to articulate their rules for ``checking out'' and ``referencing'' specific items, as well as any penalties for misuse.
    \item \textit{Derivative Works Brainstorm}: Participants were asked to consider both foreseeable and unexpected ways their images might be reused to prompt discussion on desired safeguards.
    \item \textit{Negotiating Allowed Uses}: Participants were invited to deliberate potential real-world use cases they may encounter, which was prompted through a list of different possible allowed uses and structured around three use scenarios (see Appendix: \autoref{tab:allowed-uses}).
    \item \textit{License Clause Exercise}: Participants reviewed sample clauses such as non-commercial or no-harm restrictions as concrete examples to assess whether these align with their organizational values.
\end{enumerate}
Based on governance tensions participants had shared during the interviews, each interactive component served as an opportunity for participants to apply their knowledge and to articulate potential red lines that would ultimately inform their final license. These exercises also sought to reinforce the significance of explicitly outlining acceptable uses as a core element of a governance plan.


\subsubsection{Modular License Design}
Following the data governance webinar, the lead author developed a set of modular licensing elements anchored in the Creative Commons (CC) framework. They selected two contrasting Creative Commons licenses to serve as starting ``scaffolds'' for participants: CC0 (Creative Commons Zero) and CC-BY-SA (Creative Commons At-tribution--ShareAlike). 
CC0 is a permissive license that places no restrictions on reuse or modification. This license effectively dedicates a work to the public domain, enabling maximum flexibility and openness. CC-BY-SA, by contrast, is a ``copyleft'' license. It allows others to copy, share, and remix a work, provided they (1) give proper attribution and (2) release any derivatives under the same license terms~\cite{lessig2004creative, kim2007creative}. These licenses represent the two ends of the openness spectrum: full permissiveness (CC0) versus reciprocal sharing (CC-BY-SA). These licenses were deliberately chosen because they are widely recognized and commonly used across open-source and open-content ecosystems. Further, these licenses are integrated into platforms such as GitHub, Hugging Face, and Kaggle; and are familiar to many organizations through their use in open media and education~\cite{lessig2004creative}. 
Creative Commons licenses are intentionally simple so that they generalize to many contexts. However, this property can obscure or flatten community-specific needs. To address this gap, we decomposed the Creative Commons license into a set of \textit{modular elements} to make their individual inspection and deliberation easier for each organization: 

• \textit{Attribution}: Requirements for how credit should be given to the community or organization. 

• \textit{Allowed Uses}: Specific contexts where dataset use is acceptable (e.g., education, advocacy, accessibility). 

• \textit{Restrictions}: Prohibited uses, such as commercial exploitation or harmful applications.

• \textit{Share-Alike}: Requirements for derivative works to adopt similar terms of openness. 

• \textit{Community Values}: Clauses to embed organizational norms, cultural protocols, or ethical standards.

\begin{figure}
    \centering
    \includegraphics[alt={Modular License Decision Flow: a flowchart of the license-building protocol. Left-to-right flowchart with seven main elements. After reviewing key terms and the Creative Commons framework, the participant selects a base license type. Choosing copyleft (CC BY-SA) routes through five required clause modules (attribution, allowed uses, restrictions, share-alike, community values); choosing permissive (CC0) skips them. Both paths end at a summarized license statement.}, width=1.0\linewidth]{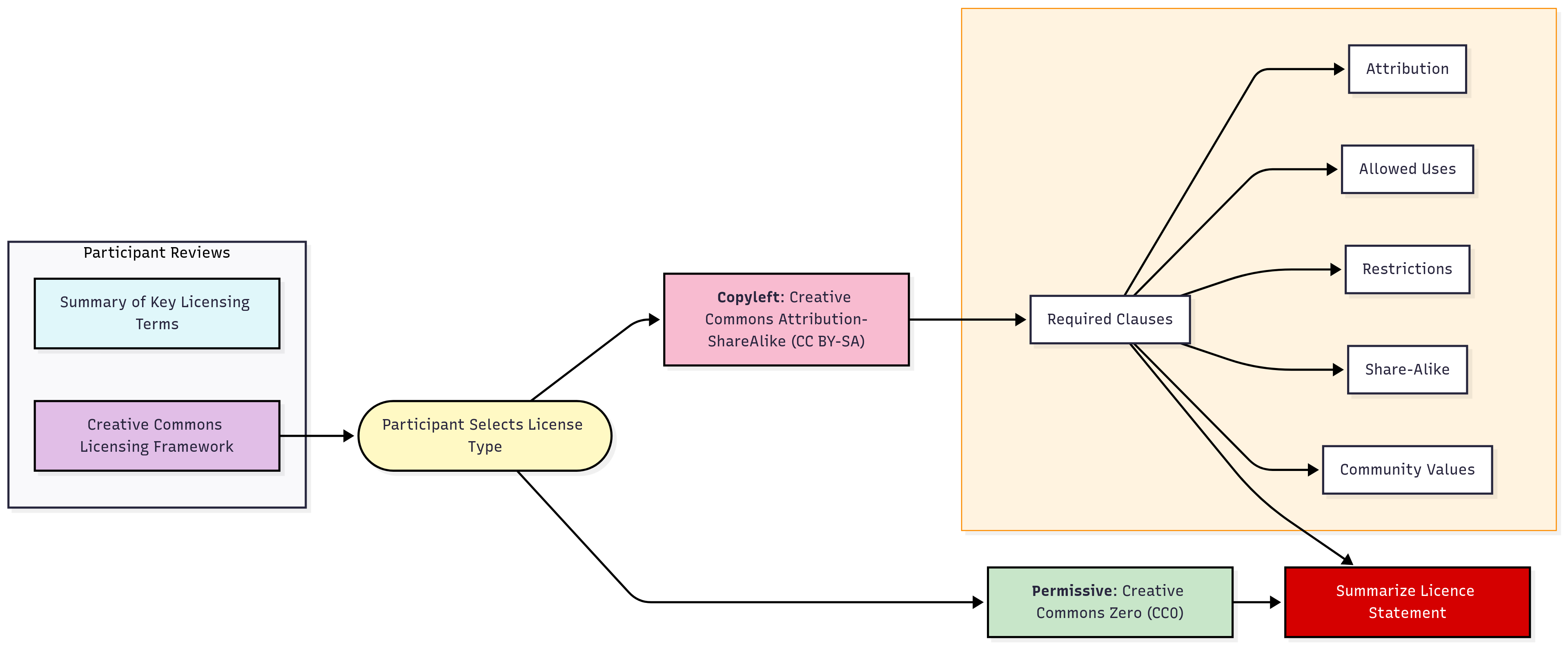}
    \caption{\textbf{Modular License Decision Flow.} Participants selected a base Creative Commons license type and layered modular clauses (Attribution, Allowed Uses, Restrictions, Share-Alike, Community Values) to produce a tailored license statement.}
    \label{fig:license-protocol}
\end{figure}

\subsubsection{Modular License Activity \& Evaluation}

\begin{figure}
    \centering
    \includegraphics[alt={The five-step license-building worksheet. A two-page fillable worksheet guiding participants from community values to a plain-language license. Parts A to C explain licensing and the Creative Commons framework; Part D selects a permissive (CC0) or copyleft (CC BY-SA) base; Part E builds the license through fill-in and checkbox fields for attribution, allowed uses and contexts, allowed modifications, restrictions, and ethical or community values; Part F drafts a plain-language summary. Full worksheet content is in the Supplementary Materials.}, width=0.95\linewidth]{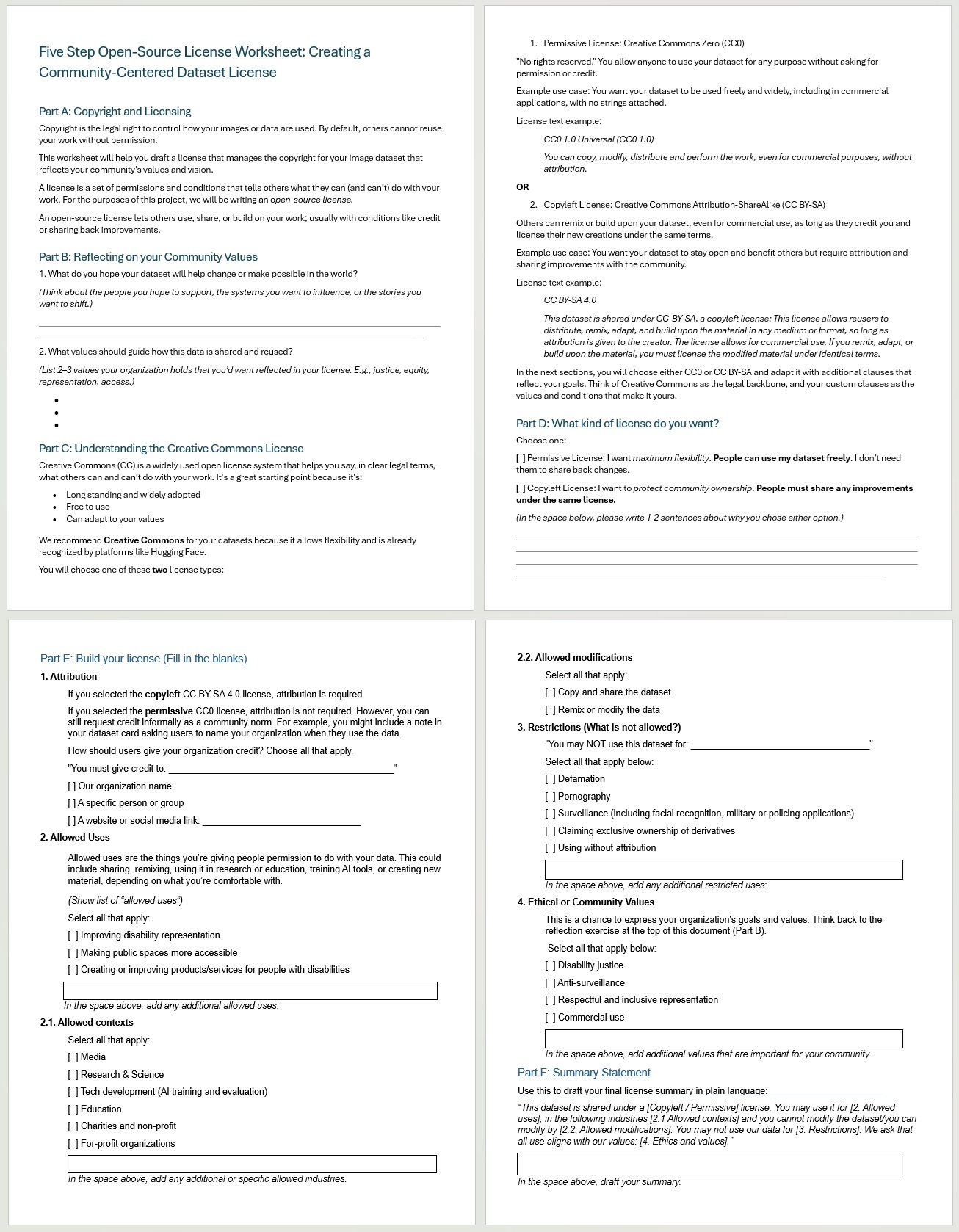}
    \caption{The five-step license-building worksheet, which guides participants from reflecting on community values through choosing and adapting a Creative Commons base license (permissive CC0 or copyleft CC BY-SA) to drafting a plain-language license summary.}
    \label{fig:module_license_worksheet}
\end{figure}

\begin{figure}
    \centering
\includegraphics[alt={Excerpt from the license-building activity showing three permission dimensions. Each option pairs an icon with a short label, and most contexts include a concrete example. 'Allowed uses' lists three purpose options; 'allowed contexts' lists six (media, research and science, tech development, education, charities and non-profit, for-profit); 'allowed modifications' lists copy-and-share and remix-or-modify. Open 'Any others?' prompts let participants add categories. Option text is given in Section 3.3.}, width=0.8\linewidth]{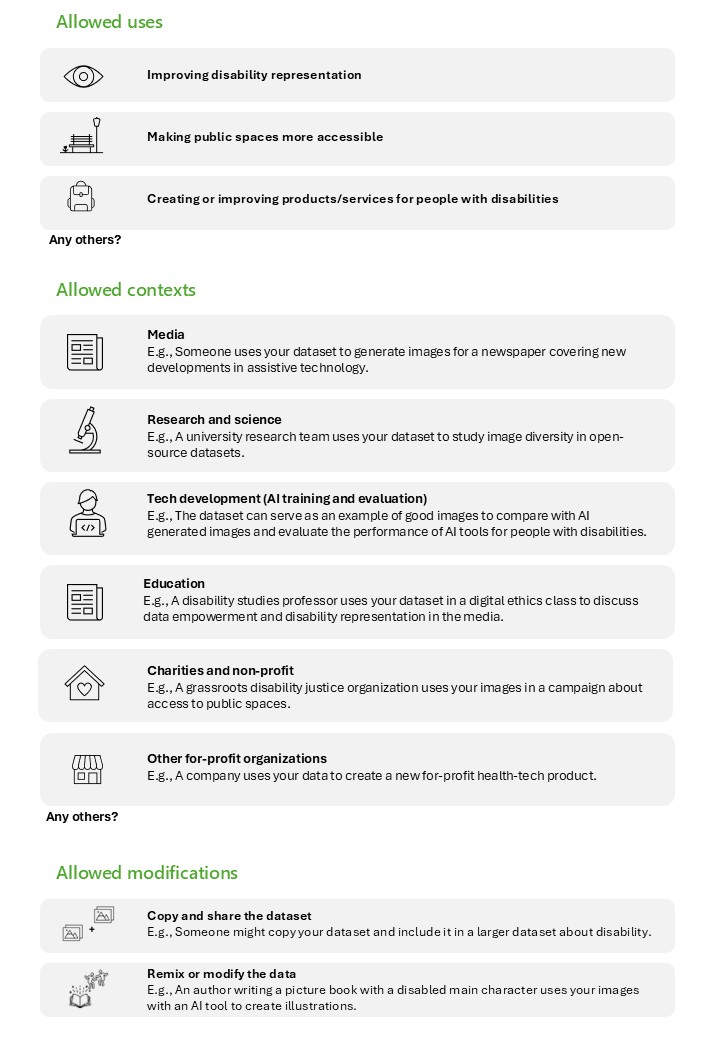}
    \caption{Excerpt from the license-building activity. Participants specified permissions across three dimensions: allowed uses (e.g., improving disability representation, making public spaces accessible), allowed contexts (e.g., media, research, tech development, education, charities, for-profit organizations), each paired with a concrete example, and allowed modifications (copying and sharing, remixing or modifying). Open-ended "Any others?" prompts invited participants to add categories beyond those listed.}
    \label{fig:alloweduses.jpg}
\end{figure}

Utilizing the new modular design as a \textit{license-building worksheet} with different elements to complete, select or expand (see \autoref{fig:module_license_worksheet}), each organization was guided through a modular license-design activity individually to observe how they work through the various parts during a 60 minute, recorded video conferencing session. 
The protocol for the activity began with a review of key licensing terms and the CC framework. Participants then selected a base license type (CC0 or CC-BY-SA) before working through each module in sequence. 
A simplified flow of the protocol is depicted in \autoref{fig:license-protocol}. For each module, participants considered example clauses and articulated whether and how these should apply to their community's dataset. To support reflection and choices for ‘allowed uses',  \autoref{fig:alloweduses.jpg} presents example prompts that both build on and extend the use scenarios introduced and discussed with project leads during data governance webinar. Furthermore, each clause was presented with contextually specific language that corresponded to governance concerns expressed by the participants during the interviews. 

These elements served as scaffolds to support participants in articulating community-specific governance boundaries. The lead researcher answered clarifying questions but, importantly, they refrained from steering choices, ensuring that participants retained autonomy over outcomes. Participants were encouraged to modify, reject, or rewrite modules entirely. The workshop culminated in a tailored license statement that integrated the chosen modules. As an initial evaluation, participants were asked to follow a think-aloud approach~\cite{charters2003use, longpre2024consent, mack2024they} to verbalize their reasoning, interpretations and in-situ reactions as they worked through the different modules, providing rich qualitative data about their sense-making process and experience. 
These accounts were supplemented with notes and observations by two additional researchers, and a 7-question survey (Appendix A: \autoref{tab:license-activity-survey}) that combines Likert-scale (e.g., strongly disagree to strongly agree) and open-ended questions to better understand the impact of the webinar and modular license design activity. 
\section{Findings}
In this section, we first elucidate the findings from the interviews (Phase 1) and the insights they provide in how high-context communities navigate formal and informal structures of governance. We then share insights derived from the modular license activity (Phase 2).

\subsection{Phase 1: Insights into High-context Communities Navigating Governance}

Our findings examine how high-context communities experience governance in practice, with the aim of offering a roadmap for thinking about data governance in communities where history, shared values, norms, and relationships shape decision-making. We use this lens to surface how formal and informal mechanisms of community structure function together in the operation of community-based organizations. In doing so, we provide context for how data licenses might need to be shaped or adapted to support community sharing of data for AI purposes. The findings are organized around three main themes: (1) \textit{organizational memory} -- describing how governance is shaped by organizational histories and the particular disability communities that organizations represent; (2) \textit{structures for decision-making, negotiation and conflict resolution} -- showing how governance operates through both formal rules or structures as well as informal mechanisms (including relationships, roles, and norms) that organize collective life; (3) \textit{membership communication and trust} -- examining the processes by which organizations communicate and maintain relationships with their members (including the tools they use and the choices available), and how these practices shape trust. We draw on examples across our three organizations to illustrate these ideas.

\subsubsection{Organizational Memory}
\label{organizationalmemory}

We found that governance across our participants was not only shaped by formal structures and written procedures, but also by organizational memory. Organizations formed their approaches through accumulated experience working with protected classes (e.g., children) and prior encounters. 

P1, an organization based in Kenya that works with people with visual impairments, for example, shared a sensitive incident involving children's data. In this instance, a third-party contractor contacted participants and families directly instead of communicating through the organization's established channels, creating confusion among parents. The episode was significant because it revealed how organizational authority and established governance relationships can become destabilized when outside actors bypass existing safeguards and procedures. This previous experience shifted how the organization approached handling sensitive participant data, requiring all members to understand ``the rights and laws, especially in the context of protected classes, such as people with disabilities and children'' (P1). Instead of being confined to their legal experts, these formalized training and oversight mechanisms became embedded in broader governance practices through compulsory participation for all members. P1 went on to describe how such organizational knowledge changed the organization's approach. They emphasized that while ``all decisions are written down'' and the organization maintains ``a handbook with clear guidelines,'' rapid technological change creates emerging ``loopholes'' where existing governance frameworks may not fully apply. In these situations, the response to prior encounters is used to adapt procedures. In the case of this organization, it is a form of institutionalized caution.

P3, an organization representing people with dwarfism with a large, heterogeneous membership, also described how organizational history shapes governance evident in their public-facing communications, where the organization’s ability to take visible positions on ``hot topics'' (e.g., the current US administration, Black Lives Matter, or COVID-19 responses) was constrained not only by its 501(c)(3) status, but also by past experiences of public messaging that provoked internal conflicts. For instance, when the organization expressed support for Black Lives Matter, individual members posted negative comments on social media or filing official complaints, leading to governance actions such as probation decisions and termination of certain product vendor relationships: \textit{“And then Black Lives Matter, that summer was gaining traction and [organization name] wanted to show support, wanted to stand firm in those current leaders wishes for our organization. And we saw a fair amount of pushback against that, of individual members speaking out on their own social media platforms or via e-mail back to our organization that they did not want to be a part of an organization that is now going to handle this. We had to cut ties with some people who sold products through our organization. So some of our members who like make products specifically for accessibility and there was 1 specific company I can think of that our organization completely cut ties with, and will no longer sell their products based on how, how they behaved after that, again, a handful of membership complaints filed. Some people were put on like probation. We've had members actually just now, five years later, beginning to be able to return from probation depending on what the decision was when their membership complaint was processed.”} (P3)

Past controversies like this shape what organizations like P3 consider feasible or too risky to communicate in the present as they balance community-building against the risks of conflict and lost (paid) membership support (see also Section~\ref{sec:communication} on communication and trust). Importantly, these experiences also influence if leaders are willing and how openly they discuss and negotiate sensitive topics with members – including the sharing of community data for AI purposes. Together, these instances illustrate how governance is shaped not only by formal rules, but also by organizational memory.

\subsubsection{Structures for Decision-Making, Negotiation, and Conflict Resolution}

Interviews across the organizations brought examples of how structures employed by organizations for decision-making, negotiation, and conflict resolution, were influenced by both formal rules or procedures as well as more informal mechanisms, such as relationships, roles, and norms. Here, P3 described their governance as a hybrid of formal, rule-based decision-making and informal, relationship-based negotiation. P3 positioned their organizations' board of directors as the primary site of binding decisions, where major issues were escalated for a vote conducted according to Robert's Rules of Order and decided through majority rule. High-stakes as decisions---such external governance information whether organizations can survey members, or whether to participate in a major external project--- are described as board-level decisions. 

While these formal structures provide the organization's official authority and accountability, we find that the practical ability to implement, contest, or reverse decisions often hinges on informal mechanisms and the uneven distribution of influence
 through cultural standing and interpersonal networks. P3 describes ``eras of change'' that follow shifts in who is elected, thereby implying that leadership style and priorities can reorient the organization even without changing formal structures. Furthermore, long-time and multi-generational members can carry informal authority (``legacies'') and ``pull extra weight,'' while committee formation may be shaped by social proximity rather than transparent criteria. Here, their organizations national conference committee is the clearest example: despite the board retaining final sign-off, membership on the committee and day-to-day control of the annual conference program operates through ``who knows who,'' which effectively creates an informal gatekeeping mechanism over one of the organization's most visible and consequential activities. During the interview, data and information practices also surfaced as a key arena where formal authority and informal autonomy can collide. For example, P3 described how the conference committee's handling of registration data (including a move to an independent website and data system not accessible to the board) created tensions about ownership, oversight, and stewardship of conference related member data. The organizational response is described as an effort to re-formalize and standardize the relationship by reasserting that the board and staff are ``legally bound'' to the organization and should therefore have clearer control over conference data processes. Yet, continuing difficulty of ``pulling'' the committee back under central ownership highlights how informal control over operational infrastructure such as systems, routines, and access can blunt formal governance, requiring careful negotiation within a ``volunteer-run'' organization, rather than pure rule enforcement.

\subsubsection{Membership Communication and Trust}\label{sec:communication}

We also see that in these community-based organizations establishing trust through member communication is in itself a governance practice. P1 described trust as something built through ``communicating clearly'' what the organization intends to do, ensuring participants understand how their information will be used, and demonstrating that organizational programs materially benefit the people they are intended to serve. They note that trust depends not only on initial consent, but also on whether participants feel included in the longer-term trajectory and impact of organizational initiatives. Communication in this context therefore functions as an ongoing relational practice through which the organization reassures participants that its governance practices are protective rather than extractive. 

Similarly, P2 described trust within their organization as emerging through interpersonal relationships among members and between members and leadership, with emphasis on information accuracy and guidance from the Constitution. When discussing how consent was obtained for collecting or sharing member data, P2 described how WhatsApp was primarily used to facilitate this process. For employment or training opportunities, for example, members were typically asked to provide consent within large group chats rather than through private messages, allowing leadership to coordinate more effectively. This communication practice reflected broader organizational norms centered around interpersonal trust and transparency, as members appeared comfortable openly sharing their information and consent with both leadership and other members within collective group spaces. 

Communication too was described by P3 as a central mechanism for managing the relationship with members. They described how the organization communicates through a mixed media ecology including emails, social media, and recurring member publications (e.g., a magazine produced three times per year), alongside the national conference as a high-audience touchpoint. Beyond information provision, these communications serve as tools that structure choices available to members: whether to participate (e.g., responding to advocacy calls, voting in elections), to seek support, or to contest organizational direction (e.g., replying to emails, commenting on posts). Here, the choice of specific channels and the framing of messaging and participation invitations (e.g., to contribute data for AI purposes) also serves to sustain a ``positive, warm, engaging community,'' while members' responses (and occasional resistance) can operate as feedback on legitimacy and representational trust. Trust here is linked to members' ability to interpret organizational intent across changing contexts and how the organization communicates during moments of external turbulence -- like news cycles or political controversies described above -- as well as internal capacity constraints.

\subsubsection{Summary of Findings}

These findings show how `collective life' within organizations is governed through both formal structures that specify where decision authority resides as well as informal, everyday relational work that is negotiated in-situ to ensure those decisions are implementable, acceptable and sustainable. In other words, formal leaders' ability to make decisions or resolve conflict needs to be aligned with informal expectations of their long-standing membership -- and their (reputational) leverage and (data) access -- to be effective. We see these informal mechanisms coming through organizational memory, the interplay between decision-making processes and relational work, and not least communication. Indeed, these findings point to `communication' as a governance practice in and of itself. Through selections of communication channels, framing of messaging, moderation of discourse as well as disclosure norms that either actively produce (or erode) members' sense that the organization represents them, listens to them, and safeguards them. The above account demonstrates how formal governance structures and practices are not static within high-context communities but interact with evolving informal mechanisms. While our findings were consistent across organizations, the ways that formal and informal practices came together differed. It is for this reason that we chose to create a modular license-building template that would allow organizations the flexibility to respond to the relational practices that were critical to their community.

\subsection{Phase 2: Insights from Community Experience of Modular License Activity}

This section integrates findings from observation of the modular license activity paired with a completion survey, drawing out the value as well as the concerns regarding community-created licenses.

\subsubsection{Translating Community Values into Data License Choices}

All participants described the modular licensing activity as a useful process of translating community values into conditions for data reuse for a dataset of 400 photos showing `good representation' of the people in their community. All three selected the copyleft Creative Commons CC BY-SA license rather than CC0, as they associated its requirements for ``attribution'' and ``share-alike'' with greater community ownership and control. While participants' survey responses indicated that the educational webinar deepened their understanding of data governance and that the modular license worksheet (\autoref{fig:module_license_worksheet}) demystified governance tools by providing a generally clear and accessible guide to governance choices, some concepts remained difficult to interpret without additional facilitation. 

P1 initially described the terminology as ``technical'' and struggled to understand the difference when distinguishing between copying, sharing, remixing, and modifying data. P2 similarly identified the Creative Commons options as the most complex part of the activity. The facilitator therefore translated legal categories into concrete, community-relevant examples before participants felt confident making a selection. P1 described the examples as ``eye-opening'' because they made it easier to explain how images could be accessed, manipulated, or remixed to people contributing data. P3 also valued comparisons between dataset sharing and the risks members already encounter when posting photographs on social media. Beyond aiding these explanations, the activity helped participants build their own fluency with licensing and legal concepts, equipping them to lead these conversations within their communities. For P3, this gave a more familiar frame for discussing risk with members who lacked technical knowledge, while also prompting concern about not wanting to ``lead my community astray.'' 
The ‘allowed contexts' module was particularly valued for its concrete examples (\autoref{fig:alloweduses.jpg}). 
Here, P2 remarked that the examples supported a more practical understanding of choices and their consequences, which helped them reinforce that their license must protect against misrepresentation while embedding cultural protocols. Likewise, P3 continuously explored -- in dialogue with the facilitator -- various imagined downstream data use scenarios to establish whether having their organization grant explicit permission could remain a prerequisite for data access, and to meaningfully think through the practical implications of data sharing in terms of how data might be used once a person ``on the other end of the computer'' has access. 

These discussions helped participants connect abstract license categories to plausible community-specific practices and consequences. The modular worksheet, supplemented by examples and clarification, subsequently enabled each project lead to modify and extend individual clauses to reflect their organization's values and political context. This was also acknowledged in the closing survey, where all participants agreed or strongly agreed that the workshop improved their understanding of data governance and how to manage data sharing and consent within their organization. 
This demonstrates how a modular licensing activity can make governance choices more accessible and values-led, yet meaningful participation depends on the effective translation between abstract legal terms and concrete, community-specific scenarios. 
Such scaffolding enables communities not merely to select a license or specific terms, but to also adapt its clauses to their cultural standards, anticipated risks, and desired forms of control.

\subsubsection{Community-Specific Concerns \& License Considerations}

Next, we describe how the modular structure enabled participants to express distinct aspirations for their image datasets, reflecting the specific priorities and concerns of their communities. While we draw on illustrative examples of individual license clauses, we intentionally do not reproduce the licenses in their entirety to preserve the confidentiality of sensitive organizational discussions. 

P1's license was guided by core values and commitment to improving representation and expanding opportunities for people with visual impairments. As acceptable uses they therefore defined: ``research and scientific development (AI and machine learning training or evaluation), accessibility improvements, making public spaces more accessible, inclusive representation efforts, [and] socially beneficial products and services.''
However, these permissions were context-dependent. On the topic of community data use in ‘media', P1 considered the use of community images in newspaper reporting potentially acceptable, yet they opposed the circulation of any such images on social media, particularly those involving children, who comprise a large proportion of their membership. This concern stemmed from prior experiences in which images of disabled people were removed from their original context and repurposed as memes or jokes, creating risks of harm that P1 had already remarked on during our previous interview. Reflecting these concerns, P1 selected for their license that data use required attribution to their organization and its online presence, permitted copying and sharing of the dataset, but prohibited remixing or modification. The final license also explicitly prohibited defamation, pornography, surveillance, claims of exclusive ownership over derivatives, and unattributed use. In the survey, P1 emphasized that those choices reflected their organization's commitment to children's rights, respectful disability representation, and compliance with the Kenyan Data Protection Act, expressing the hope that users would handle the dataset ``responsibly and in line with our commitment to protect vulnerable groups.''
Overall, this suggests that for P1 and their organization, openness to data sharing was conditional based on terms that ensured safeguarding of children, that prevent degrading portrayals of disability, and that helped preserve the intended context of representation. 

For P2, the primary goal of the license was to ensure that any reuse of the dataset aligned with community values of inclusion, diversity, and awareness, and contributed to ``a positive reflection of who we are in the community.'' They viewed the image dataset as a means of demonstrating ``how we want ourselves to be represented,'' rather than leaving those decisions to external actors or AI systems. Consequently, P2 was comfortable permitting use across many contexts including: media, research, education, AI development, and, under appropriate conditions, was considering commercial use. To better understand the implications of these permissions, P2 requested additional examples of how companies, charities, and nonprofit organizations might use their organizations data. On the back of these and other examples, P2 chose to clearly prohibit data uses related to defamation, pornography, surveillance applications (including facial recognition and military or policing uses), stigmatizing content, discriminatory profiling, privacy violations, and re-identification. Further, their license required that all uses actively support ethical and organizational values of ``disability justice, anti-surveillance, stigma reduction, equity, diverse and respectful representation and non-exploitative practices''. In other words, for P2, the central governance challenge was not limiting data use, but ensuring that any reuse remained aligned with community values. Openness to data sharing was acceptable when accompanied by explicit protections against harmful, exploitative, or decontextualized uses, and an active commitment to respectful representation. 

Lastly, P3 sought fair and appropriate inclusion in disability and diversity representation -- while naming empowerment and resilience as central values within their license. As discussed in the section on ‘organizational memory', their licensing decisions were strongly shaped by a history of harm associated with the circulation of images of people with dwarfism. Consequently, the license prohibited the use of their image data in media and by for-profit organizations. Most significantly, P3 included a use restriction against entire industry sectors pertaining to (1) pharmaceutical use and (2) entertainment. These restrictions reflected longstanding organizational opposition to requests for community materials to be used in marketing corrective treatments or medications for people with dwarfism to which their organization had ``always stood firm'' against. To P3, the definition of these restrictions reflect an affirmative understanding of dwarfism that emphasizes how community members are proud of their identity and did not wish to be represented as needing to be ``altered,'' ``changed,'' or made ``less disabled''. Consequently, the license explicitly states that ``it is vital that this dataset is not used in any pharmaceutical setting (i.e. marketing, medical research, treatments, etc.) or entertainment setting (including but not limited to wrestling, circus performance, or costumed characters that perpetuate negative stereotypes of dwarfism)''. The closing section of their license further reinforces this position and organizational values: ``we embrace our individuality and uniqueness. We believe disability is an essential pillar of human diversity, and we do not wish to be altered or changed.'' Thus, for P3, the license became a way to express community values not captured by generic categories: an explicit commitment to disability pride, embracing individuality, and rejecting cure-oriented interpretations of disability.

\subsubsection{Points of Uncertainty \& Limitations of Licensing as Governance Safeguard}

Despite the flexibility of the modular licence-building worksheet, participants raised three main concerns: (1) whether license terms were sufficiently specific, (2) whether downstream users would comply with them, and (3) how licenses would operate across different legal jurisdictions. P1 worried that broad categories in the license could either permit harmful uses or unintentionally exclude beneficial ones, noting that there was ``so much to break down and understand before picking.'' Most notable were concerns about enforcement. P1 asked whether compliance could be tracked or whether the terms were ``just suggestions in the hope'' that users would respect them, or if remedies would exist in cases of data misuse, e.g.: ``Does monitoring for compliance fall on the original owner of the data? Does Hugging Face take responsibility for monitoring use?'' (P1). Similarly, P3 expressed concern that images of people with dwarfism could still be used ``inappropriately or without consent,'' asking who would be responsible for monitoring and enforcing license terms. Here, discussions with the lead researcher around sharing datasets on platforms such as Hugging Face provided some reassurance, as licenses are a familiar and established mechanism for governing dataset use in these contexts: ``I know like nothing can be perfectly assumed, promised, you know, whatever at the end of the day because it's the Internet. But is it fair to say the people using hugging face are already familiar players with licenses. They're already familiar with like the protections of data. This isn't just the average person off the street?'' (P3). 

Still, participants recognized that licenses cannot fully prevent misuse. Participants also raised governance concerns that extended beyond the scope of licensing itself. These included consent procedures with community members; maintaining contextual integrity; identifying realistic control mechanisms that resource-constrained (non-profit) organizations could implement; as well as addressing issues such as cross-border data transfers, applicable laws, and secure data archiving or deletion. Such concerns exceed what a modular copyright-focused Creative Commons license could readily address. Thus, although our license activity helped participants make their data governance values and boundaries explicit, it does not itself provide the technical controls, monitoring arrangements, life-cycle policies, or enforcement resources needed to guarantee those boundaries. This positions community-led licensing as only one component of participatory data governance rather than a complete safeguard against unintended use. Ensuring responsible data use also requires complementary technical, organizational, and legal mechanisms that support accountability and stewardship across the data life-cycle.

\subsubsection{Summary of Findings}
Our findings exhibit a dual reality. Participatory governance interventions illustrate how disability organizations can adapt Creative Commons scaffolds into highly contextual instruments. The licenses represent both convergence (a shared legal foundation) and divergence (tailored permitted uses and restrictions, and value statements) in how high-context communities can govern their data. The CC framework is low in complexity, and CC license clauses can be adapted to articulate community specific values and political red lines. A shift from rigid legal protocols, such as consent, to concrete and community-oriented governance clauses was one of the most consistent outcomes. At the same time, there are persistent challenges such as the abstruseness of legal language, and the lack of enforceability. In this sense, participatory approaches can build capacity and generate licenses that communities recognize as their own, but they also reveal where additional external supports may be needed. Licensing is not a standalone solution, but part of a broader governance ecosystem that must account for stakeholder negotiation and the ongoing work of organizational care.
\section{Discussion}
In this section, we reflect on the broader significance of our findings and situate them within ongoing debates around participation in data governance and AI licensing~\cite{lin2026, lin2025, joon2026, policycraft, hudig2025}. We organize our discussion in three parts. First, we use our findings to outline design considerations for governance tools emerging from the layered, relational tensions that disability organizations face (RQ1). Second, we position licensing among other governance mechanisms, arguing that a modular, low-complexity licensing approach offers a tractable entry point for high-context communities, while acknowledging its limits (RQ2). Finally, we outline implications for platforms and practitioners who seek to scaffold community-led data stewardship in practice.

\subsection{Translating Governance Tensions into License Design Considerations}
Our analysis demonstrates how disability organizations experience governance as an ongoing negotiation of tensions and trade-offs, not a fixed set of rules~\cite{jazette2026}. Governance is layered, and combines formal policies and informal practices (such as ad hoc responses). Governance is simultaneously deeply relational, grounded in trust and contextual histories of harm. We interpret these tensions as three connected design considerations for any governance tool serving high-context communities: (1) procedural clarity; (2) relational accountability, and (3) reflective learning. 

\subsubsection{Procedural Clarity} 
Across each organization, leaders grappled with determining who holds authority over data decisions, how consent should be understood in collective contexts, and what contextual information must accompany a dataset to ensure responsible use. When roles or processes were ambiguous, such as third parties contacting families directly, or images circulating beyond expected contexts, our participants described feeling legally exposed and ethically constrained. This resonates with critiques of conventional governance frameworks which overemphasize individual consent while under-specifying institutional responsibilities~\cite{longpre2024data,simms,young2024participation}. Within high-context communities, governance tools must make decision points explicit. This means articulating who is consulted, under what conditions, and with what documentation~\cite{young2024participation}. In our work, this requirement surfaced as concrete license clauses through which our participants clearly specified ‘allowed uses', and sector-specific exclusions. 

\subsubsection{Relational Accountability.} Participants framed many of their most difficult decisions not as technical or legal issues, but as questions of trust. Breakdowns in trust after conflicts over political messaging, or perceived overreach by external actors had lasting effects on member participation and organizational cohesion. 
P3 illustrated this vividly. With a large, heterogeneous membership, they described how public-facing communication is rarely read as neutral, since any statement can be taken as ``speaking on behalf of the entire membership'' even when members' views diverge. Constraints on taking visible positions on ``hot topics'' were attributed both to their 501(c)(3) status, framed as needing to ``keep politics out of our organization,'' and to hard-won experience. Past advocacy, such as expressing support for Black Lives Matter, drew negative member responses and formal complaints, leading to governance actions like probation decisions and severing ties with vendors. These breakdowns create acute friction and tension.

Prior work shows how relational approaches to data governance can help bolster trust by centering the relationships through which data is produced~\cite{viljoen2021relational}. Along these lines, our study shows that governance tools must encode whose interests are prioritized. In licensing terms, this appeared in strong no-harm clauses and bans on specific sectors (e.g., pharmaceutical, military, entertainment). Our participants also constructed community specific value statements that articulated their commitments (e.g., empowerment through advocacy, safeguarding children). 

\subsubsection{Reflective Learning.}
Organizations described challenges in updating existing rules to fit new technical contexts. Policies written for in-person events did not translate cleanly to platform-based data sharing, and national data laws offered limited guidance for emerging practices such as AI-generated media. These gaps produced tensions where formal rules and everyday practices diverged, aligning with critiques that static compliance frameworks can result in organizational harms~\cite{bogen}. Adaptable governance tools that support ongoing reflection and revision make it possible for communities to revisit earlier decisions and adjust terms as their needs evolve~\cite{choksi2024licenses,coleman2005three}. 
These design considerations are not exhaustive. They are a starting point to understand how high-context communities experience governance in practice, and what they need from governance tools. In our study, these considerations bridged real-world governance struggles and concrete licensing decisions.

\subsection{Licensing as a Community Data Governance Mechanism}
Licensing tools sit alongside other governance models, such as data trusts and cooperatives. Prior work has called for intermediary institutions that can hold and govern data on behalf of communities~\cite{botero2022even,delacroix2019bottom}. While these models foreground collective control, they often assume access to robust legal infrastructures, organizational capacity, and high levels of data rights literacy~\cite{olorunju2024african}. Our partner organizations described operating under tight resource constraints, and limited in-house legal expertise. In this context, licensing emerged not as the ideal governance solution, but as a tractable governance tool they can meaningfully shape within existing constraints. For example, a license can be attached to a dataset hosted on a platform like Hugging Face. It can be read and discussed by members, and it can provide a basis, however imperfect, for contesting harmful reuse. 

At the same time, we underscore the limitations of relying on licenses alone. Our participants voiced concerns about enforcement and monitoring, and cross-border data flows. Their concerns echo critiques of ethical AI licenses that question their effectiveness against actors who may simply ignore them~\cite{choksi2025brief,contractor2022behavioral,longpre2024consent}. For disability organizations with histories of misrepresentation and exploitation, these unresolved questions shape whether open licensing feels viable at all. To remedy these concerns, we present licensing as one governance mechanism within a broader ecosystem that should also include institutional and platform supports. 
We show that open-source licenses are \textit{a} tool that offers communities a rare degree of control. Their value lies not in solving every governance challenge, but in enabling low-resource, high-context communities to articulate and enforce -- to the extent possible -- the terms under which their data should circulate.

\subsection{The Modular License as an Adaptable Tool}
A central contribution of this paper is the modular license design. Using the accessible structure of low-complexity copyright frameworks (i.e. Creative Commons), we decomposed licensing into a series of structured decision points. The protocol, depicted in Figure 1, treats licensing as a participatory activity rather than a purely legal exercise. Aligning with calls in HCI and AI governance for participatory, justice-oriented approaches that meet communities where they are~\cite{sociotechnicalaigov,kamikubo}, our tool lowers both the informational and cognitive burdens associated with complex AI copyright regimes. Some Responsible AI Licenses ask organizations to parse dense, pre-written legal documents~\cite{contractor2022behavioral}. The modular approach instead invites non-experts to work through a license one clause at a time, in plain language, grounded in their own histories and priorities. This makes relational governance visible and negotiable. For example, our participants debated whether sectors like pharmaceuticals or entertainment should be categorically banned, or whether commercial use might be acceptable only when it demonstrably benefits disabled people. They negotiated implicit community norms about reciprocity that would otherwise remain unspoken. As such, the modular license functions as a communal boundary artifact~\cite{choksi2024licenses,kelty2020two} around which community members, researchers, and potentially platforms can reason about data governance across diverse contexts. Finally, the modular structure of our protocol is inherently adaptable. Although instantiated around Creative Commons licenses in this study, it can be extended to other low-complexity licenses (e.g., MIT, Apache) or paired with institutional models~\cite{delacroix2019bottom,olorunju2024african}. Clauses can be added, revised, or retired as communities gain experience or as external conditions change. In this sense, the modular license protocol offers a reusable format for co-designing governance tools with communities.

\subsection{Implications for Platforms and Practitioners}
We propose three concrete steps that platforms and practitioners can take to support participatory AI data governance for high-context communities. 

\paragraph{1. Guided Modular License Builders.}
Platforms that host datasets and models often present license options in the form of a list to select and apply predetermined options. 
This design reproduces a familiar limitation of consent-based governance, where predetermined options are accepted without the deliberation needed to make them meaningful~\cite{longpre2024consent}, and it is poorly aligned with the needs of high-context communities.
Such communities require space to deliberate and articulate context-specific conditions. We recommend that platforms offer guided license-building interfaces that start from simple base licenses and walk users through modular decision points (i.e. attribution, allowed uses, restrictions, sector exclusions, value statements), using plain-language prompts and examples. This would operationalize the structured reflection we piloted in our workshops at platform scale. 

\paragraph{2. Embedding Participatory Education.}
Participants stressed that much of the value of our intervention lay in the educational aspects. Specifically, understanding basic licensing concepts, platform affordances, deliberating scenarios of potential misuse, and connecting these back to their own governance histories. 
This responds to a known limitation of consent-based governance, where the burden of understanding complex terms falls on individuals who may lack the resources or expertise to interpret them~\cite{RENAUD2018198, simms, nissenbaum2011contextual}. Building that comprehension is itself a governance intervention. Documentation and license selection are often treated as purely administrative steps, but platforms can instead embed lightweight educational materials, scenario-based prompts, or even facilitated office hours into data onboarding workflows. Interactive resources would not eliminate the need for legal advice, but would help communities interpret and adapt governance tools.

\paragraph{3. Additional Governance Infrastructures.} 
As governance artifacts, licenses articulate community norms but cannot on their own resolve questions of enforcement, monitoring, or redress~\cite{choksi2024licenses}. Platforms play a critical role in building infrastructures that allow communities to request takedowns, and document misuse or patterns of harmful behavior. Additional mechanisms for attaching community-authored ‘terms of use', logging when and how data is accessed, or supporting collaborations with intermediary organizations that can help enforce community-defined conditions. These steps would move platforms away from being neutral hosts and toward active partners in relational governance~\cite{sociotechnicalaigov, suresh2024participation, young2024participation}.

\section{Limitations and Future Work}
Our study raises several important limitations that shape the scope and interpretation of our findings. First, our work engaged three disability organizations. While this research design afforded depth, it necessarily limits the breadth of perspectives captured. Future work could include broader community deliberation and direct observation of organizational decision-making. Our findings are shaped by the specific disability groups and cultural contexts represented in this study. Histories of harm may differ significantly across disability (and other marginalized communities), as well as across international geopolitical and legal contexts. Our results should therefore be understood as transferable rather than generalizable. 

As researchers, we occupied dual roles as facilitators and educators. Despite our efforts to minimize influence and preserve participant autonomy, our presence may have subtly shaped how participants reasoned about licensing. 
Future work should experiment with approaches that further distribute power, including peer-led facilitation models. Our study did not include longitudinal evaluation of how the community licenses are adapted over time. Future work could examine the real-world operationalization of community-crafted licenses and their uptake on platforms. Finally, we consider the opportunities and risks of emerging AI-assisted tools for governance. While generative AI may help lower barriers to drafting or modifying licenses, these systems may also introduce inaccuracies or obscure the legal and political stakes of governance decisions. Future research should also explore how AI systems can be designed to support, rather than replace, community deliberation---ensuring that automation does not reproduce the very forms of disempowerment that participatory governance seeks to remedy.

\section{Conclusion}
Our research demonstrates that contextually-aligned license design can enable disability advocacy organizations to engage more safely and meaningfully with AI development.
Working with three disability advocacy organizations, we described how organizational memory, a mix of formal and informal relational practices, communication, and the trust these produce, jointly shape how communities evaluate the opportunities and risks of sharing (image) data for AI.
Our modular license activity translated these community norms and situated governance tensions into concrete choices about attribution clauses, permitted uses, restrictions, safeguarding, and community values, enabling each organization to reflect relational governance needs that standardized approaches and tools overlook.
Yet, licensing cannot guarantee appropriate use: monitoring, enforcement, consent, and cross-border accountability require complementary technical, organizational, and legal infrastructures.
For HCI, our findings position license design as a participation activity and a practical entry point into broader AI governance that prioritizes shared stewardship and contextual sensitivity.
Going forward, we argue that designing platforms that support guided deliberation, community-authored terms, revision, and redress can help shift AI data stewardship from standardized permission-setting toward ongoing, context-sensitive relations of care.

\bibliographystyle{ACM-Reference-Format}
\bibliography{bibliography}
\clearpage
\appendix
\setcounter{secnumdepth}{0}
\section{Appendix}
\label{appendix:A}

\begin{table}[h]
\centering
\small
\caption{Interview Protocol: Four data governance themes with sample interview questions and corresponding participant responses.}
\label{appendix:interview-protocol}
\begin{tabularx}{\linewidth}{>
{\raggedright\arraybackslash}p{3.5cm} >
{\raggedright\arraybackslash}p{5.4cm} >
{\raggedright\arraybackslash}p{5.4cm}
}

\toprule
\textbf{Governance Theme} &
\textbf{Sample Interview Question} &
\textbf{Participant Responses}\\
\midrule

Organizational Values and Governance &
\textbullet\ When you think about how your organization represents your community to the broader public, what principles do you strive to center?
&
P1:``\textit{Positive representation... dwarfism isn't something to fear, [it's] a human experience}.'' \\
& \textbullet\ How are decisions about data sharing and technology typically made in your organization?
& P3:``\textit{A department which governs data... they make the overall decisions of what is shared}.'' \\
\midrule
Data Practices and Consent
& \textbullet\ How does your organization approach consent when it comes to collecting or sharing member data?
& P2: ``\textit{I will probably share on the WhatsApp group... when people respond with 'consent', it is typically in the group so others can see}.'' \\
& \textbullet\ What expectations do you think your community members have around being asked before their information is used in new ways? 
& P3:``\textit{[Whether] their information is being used to benefit them in a direct way...[and] the end result of the project is shared back with them}.'' \\
\midrule
Conflicts, Tensions, and Organizational Learning
& \textbullet\ Has the organization faced situations where information sharing has caused tensions or concerns?
& P3:``\textit{There was a concern: how do you handle data leaks, especially if it's a third party handling our data?}'' \\
& \textbullet\ Were there ever moments when you felt your organization's values were in tension with a funder, partner, or external party?
& P1:``\textit{We had to cut ties with specific vendors based on their stance on [political movements]}.'' \\
\midrule
Roles and Communication
& \textbullet\ In your organization, are there people with explicitly defined roles? If so, can you identify these roles?
& P2:``\textit{Yes, many. Chairperson or president, secretary, treasurer, volunteers}.'' \\
& \textbullet\ Are there times when you want to explicitly communicate with members in a specific role? If so, please describe a scenario when that occurred.
& P1:``\textit{I do reach out to specific committee members explicitly and ask for their expertise... typically to help someone with a specific need}.'' \\
\bottomrule
\end{tabularx}
\end{table}

\begin{table}[b!]
\centering
\caption{Sample ``Allowed Uses'' scenarios presented during the participatory webinar. These scenarios helped organizations deliberate acceptable uses and identify red lines.}
\label{tab:allowed-uses}
\begin{tabularx}{\linewidth}{@{}lL@{}}
\toprule
\textbf{Category} & \textbf{Example Scenarios} \\
\midrule
Purpose-driven uses &
\textbullet\ Improving disability representation in media.\\
& \textbullet\ Making public spaces more accessible.\\
& \textbullet\ Developing products and services for people with disabilities.\\
\addlinespace
Context-specific uses &
\textbullet\ A university research team analyzes dataset diversity.\\
& \textbullet\ A disability studies professor uses images in a digital ethics class.\\
& \textbullet\ A non-profit runs a public campaign on access to public spaces.\\
& \textbullet\ A company creates a for-profit assistive technology product.\\
\addlinespace
Modification-based uses &
\textbullet\ Copying the dataset into a larger dataset on disability.\\
& \textbullet\ Using images as a base for AI-generated illustrations in educational materials.\\
\bottomrule
\end{tabularx}
\end{table}
\clearpage

\begin{table*}[t]
\centering
\caption{License Activity Survey}
\label{tab:license-activity-survey}
\small
\begin{tabularx}{\textwidth}{p{0.05\textwidth}X}
\hline
\textbf{No.} & \textbf{Survey item and response options} \\
\hline

1 & The license workshop helped me better understand how to manage data sharing and consent in my organization.\\
  & $\square$ \textit{Strongly disagree }\quad
    $\square$ \textit{Disagree} \quad
    $\square$ \textit{Unsure} \quad
    $\square$ \textit{Agree} \quad
    $\square$ \textit{Strongly agree} \\
\hline

2 & What part of the worksheet sparked the most discussion or felt most difficult to answer? \\
& {\textit{[Open text response]}} \\
\hline

3 & Did any past concerns (e.g., about misuse, power dynamics, or partner pressure) arise again during the licensing process? \\
& {\textit{[Open text response]}} \\
\hline

4 & Do you feel that the licence you created reflects your community's values concerning data sharing and representation? Why or why not? \\
& {\textit{[Open text response]}} \\
\hline

5 & I feel more confident about our organisation's role as a data steward after completing this activity.\\
  & $\square$ \textit{Strongly disagree} \quad
    $\square$ \textit{Disagree} \quad
    $\square$ \textit{Unsure} \quad
    $\square$ \textit{Agree} \quad
    $\square$ \textit{Strongly agree }\\
\hline

6 & Is there anything you remain unsure about or are still considering regarding data governance and licensing? \\

& {\textit{[Open text response]}} \\
\hline

7 & If another organisation asked what you learned about sharing data responsibly, what would you tell them? \\
& {\textit{[Open text response]}} \\
\hline
\end{tabularx}
\end{table*}


\end{document}